\documentclass[titlepage]{amsart}
\usepackage{amsaddr}
 \usepackage{color, soul}  
\usepackage{ulem}                                         
\usepackage{amsfonts}
\usepackage{amsmath}
\usepackage{setspace}
\usepackage{graphicx}
\usepackage{psfrag}
\usepackage{graphics}
\usepackage{amssymb}
\usepackage{bm}
\usepackage{multicol}
\usepackage{amsmath, amssymb,mathabx,mathrsfs}
\usepackage[toc,page]{appendix}
\usepackage{comment}
\usepackage{epsfig}
\usepackage{graphicx, subfigure}

\theoremstyle{definition}

\theoremstyle{remark}

\numberwithin{equation}{section}

\theoremstyle{summary}

\begin{document}
\title[ ]{Permanent Capture into the Solar System }

\author{EDWARD BELBRUNO}
\address{Yeshiva University, Department of Mathematical Sciences, New York, NY 10016, USA }
\email{edward.belbruno@yu.edu  (corresponding author)}

\author{JAMES GREEN}
\address{Space Science Endeavors, Silver Spring, Maryland 20901 USA}
\email{jlgreen1@earthlink.netd}

\begin{abstract}
In this note, an interesting region about the Sun in phase space is described where the permanent capture of an object, $P$, of small mass  from interstellar space can occur, under the gravitational perturbation of the resultant mass of the Galaxy. $P$ is never ejected back into interstellar space and won't collide with the Sun. It cycles about the Sun for all time, asymptotically approaching the capture set.  In addition to being permanently captured, $P$ is also weakly captured.   A recent result in \cite{Belbruno:2024} describes this region in general. It has a fractal structure.  A capture cross section and the probability of permanent weak capture are estimated.  This is applied to the permanent capture of rogue planets and other objects from interstellar space
\end{abstract}

\date{March 30, 2024}

\maketitle

\section{Introduction}  \label{sec:Intro} 
\medskip

The permanent capture of a small body, $P$, about the Sun, $S$,  from interstellar space occurs when $P$ can never escape back into interstellar space and remains captured within the Solar System for all future time, moving without collision with the Sun.  The perturbing gravitational field is the resultant gravititational field of the Galaxy due to dark and baryonic matter, as estimated in \cite{BelbrunoGreen:2022}. The gravitational field of the Galaxy as a tidal force in the Solar System, modeling dark matter,  is previously studied in \cite{HeislerTremaine:1986}, \cite{Levinson:2001}, \cite{Levinson:2010}, \cite{Nesvorny:2017} and in other studies.

The capture of comets and other objects into our Solar System has been previously studied in a number of papers; for example, \cite{Napier1:2021}, \cite{Napier2:2021}, \cite{Hands:2020} (see also \cite{Heggie:1975}, \cite{LiAdams:2016}, \cite{LaughlinAdams:2000}).
 The focus of these papers is generally to understand long term behavior of captured bodies, from an astronomical perspective where by 'permanent capture' they mean a finite time span. For example, in  \cite{Napier1:2021} this is $1$ billion years. The general methodology \cite{Napier1:2021, Napier2:2021, Hands:2020} use is to perform a Monte Carlo analysis for many trajectories and estimate capture cross sections. It gives a sense of the longevity of capture for a sampling of the phase space for only a finite time.  It is also noted that these papers don't model the gravitational force of the Galaxy, but rather only model the gravity of Jupiter as the third body perturbation. This is different modeling from what we are assuming in this paper. The distance $P$ is moving in this study is far from $S$, at $3.8$ LY  (see Section \ref{subsec:InitialAssumptions}), where the gravity of $S$ represents the gravitational force of the Solar System, and the third body is represented by the gravity of the Galaxy as a tidal force. This tidal force has an appreciable effect on the structure of the phase space for the velocity range and distance from the Sun we are considering.  

The focus of this paper is more theoretical in nature, which is on studying the dynamical and topological properties of a special type of permanent capture, called permanent weak capture which occurs for infinite time. The body never escapes after capture.  A simplified modeling is assumed, and the goal is to describe a dynamical mechanism for permanent capture and how it works.  Its extension to a more realistic modeling is not described, although it is conjectured that permanent capture regions would still exist. Probabilities of permanent weak capture are estimated and capture cross sections estimated.

A dynamical mechanism is described for the permanent capture of $P$ from interstellar space about $S$,  making use of recent results in \cite{Belbruno:2024} (since this reference is cited frequently, it is denoted by $B24$).  These results show that when $P$ is on a permanent capture trajectory from interstellar space, it also is approaching a special weak capture set about $S$, in phase space,  as time , $t \rightarrow \infty$.  This is where the Kepler energy between $P$ and $S$ becomes negative. 
Taken together, this is permanent weak capture.    The permanent weak capture set is an interesting fractal set of zero measure in phase space.

To show how this can occur, a simplified modeling is assumed. It assumes three point masses: $P$, $S$, and $P_{MW}$, which has the virial mass of the Galaxy placed at $CG$, the center of the Galaxy. $S$ is assumed to move about $P_{MW}$ in a circular orbit. $P$ has a negligble mass.  This defines the three-dimensional restricted three-body problem. 

The definition of permanent capture used here is more restrictive that what is used in the classical definiton by J. Chazy for the Newtonian three-dimensional three-body problem for arbitrary mass points $P_0, P_1, P_2$ (\cite{Chazy:1918}). In that case $P_0$ is permanently captured by $P_1, P_2$ if $P_0$ approaches $P_1,P_2$ from infinitely far away, where for all future time, $P_0$ remains a finite distance from $P_1,P_2$, and where $P_1, P_2$ stay a bounded distance from each other and no bodies collide. It is shown in \cite{Chazy:1918} that the set of permanent capture trajectories in the general three-body problem is a small set of measure zero. Showing it actually existed for $P_i, 1=0,1,2$ was an important conjecture, proven many years later by V. Alekseev for the Sitnikov problem, using properties of chaotic dynamics   (\cite{Alekseev:1969}) .   

In this note, $P_0= P$, $P_1 = P_{MW}$, $P_2=S$.   
We consider a special case of Chazy's definition where the permanent capture of $P$ is about $S$,  and $P$ does not approach $S$ from infinitely far away, but rather from a finite distance. Also, the requirement of $P$ remaining a bounded distance from $S$ is replaced by $P$ remaining within the Solar System.  The approximate extent of the  Solar System is estimated to be $3.81$ LY  in radius from $S$ (see Section \ref{sec:RestrictedProblem}). It is also shown that if dark matter were not modeled, then the the Solar System would extend out much further.

In numerous numerical studies for two general mass points, $P_1$, $P_2$, moving about their common center of mass in circular orbits, where $P_1$ is much more massive than $P_2$, temporary weak capture of $P_0$ about $P_2$ is typically considered. Temporary weak capture has been used in practice to find low energy trajectories from the Earth to the Moon where spacecraft are automatically captured about the Moon \cite{Belbruno:2004, BelbrunoMiller:1993}.

To show that permanent capture exists, a planar motion is considered, where $P$ moves in the same plane as the circular motion of $P_1, P_2$. This is done since the rigorous results and formulation for weak capture are for the planar case. This is sufficient for this Note since its goal is to show dynamically how permanent weak capture can occur. A similar process would be expected in the three-dimensional case, but this is beyond the scope of this note.  Understanding these results in a more realistic setting is also beyond the scope of this note.

A region about $P_2$ in the phase space of position and velocity coordinates is first estimated by monitoring the stability of cycling trajectories of $P$ about $P_2$ after $n$ cycles, $n = 1,2, \ldots$.  It defines points where there are transitions between stable and unstable cycling motion for trajectories and starting with negative Kepler energy relative to $P_2$ (\cite{Belbruno:2004, GarciaGomez:2007, BelbrunoGideaTopputo:2010}). This region is called the $n$th weak stability boundary, labeled $W_n$. It consists of points giving unstable cycling for trajectories about $P_2$. By symmetry, this yields weak capture trajectories to points on $W_n$ after $n$ cycles. Although this region is rigorously defined in the planar three-body problem, it was initially defined in the general three-dimensional three-body problem in 1987 \cite{Belbruno:2004}.

The structure of $W_n$ about $P_2$  has been an open problem.   It was conjectured to be fractal in nature (\cite{GarciaGomez:2007}) (GG07 for brevity) and also to give rise to chaotic motion for any trajectories taking initial values on it.    Recent results in $B24$ has determined the structure of this region in the case $n \rightarrow \infty$, that is, for infinitely many cycles. It is defined by the set $W' = \lim{n \rightarrow \infty} W_n$.    $W'$ turns out to be fractal set with a Cantor set structure.   Moreover, $W'$ bounds a region about $P_2$  where only infinite stable cycling motion can occur for all time. This region shares similar properties to the classical Mandelbrot set for a simple complex quadratic map'

In this note it is shown that permanent weak capture of $P$ occurs on a special subset of $W'$ (Section \ref{sec:PermCap}).   This is applied to the case where $P_2 = S$ and $P_1 = P_{MW}$.  The gravitational forces acting on $P$ is $F_S$, the gravitational force of the Sun, and $F_{G}$, the resultant gravitational field due to $P_{MW}$ relative to the CG. This is a tidal force relative to $S$.  The permanent capture of $P$ occurs from interstellar space where $P$ first moves through a channel into the Hill's region about $P_2$ at $3.81$ LY from $S$, and then performs infinitely many cycles about $S$ to asymptotically approach weak capture about $S$ at one of the special points of $W'$.  $P$ can never escape back into interstellar space.  

A capture cross section is estimated in Section \ref{sec:PermCap}.  It is shown to have a substantial area, but not all the points on it yield weak permanent capture trajectories. The points on it that yield weak permanent capture trajectories form a fractal set of fractional dimension. This is valid for a restricted range of the Jacobi constant.  The probabilities of permanent weak capture are determined. A range of the probabilities are given in (\ref{eq:probrange}). 

The capture of rogue planets is discussed in Section \ref{subsec:RoguePlanets}. The gravity of such objects traveling within the inner solar system could perturb the orbits of the planets. It is possible such a perturbation could be detected.  

The permanent capture of Rogue planets is an important application since they can accumulate over time. This implies that even though the permanent capture set is small and therfore capture of a rogue planet is unlikely, over time on the order of hundreds of millions of years, the liklihood of capture increases. 

The analytic model for $F_G$  is described in Section \ref{sec:NewtonianModeling}.  This, together with $F_S$ and the centripetal force, models the motion of $P$ with a classical set of differential equations, the restricted three-body problem (see Section \ref{sec:RestrictedProblem}).  $W'$ and permanent weak capture is described in Section \ref{sec:PermCap}.  Capture cross sections are computed, and a probability of permanent weak capture is estimated in this section.

\section{Galactic Force Modeled as a Classical Point-Mass Inverse Square Newtonian Force} 
\label{sec:NewtonianModeling}

 \noindent
   In \cite{BelbrunoGreen:2022} ($BG22$ for brevity)  ${\bf F_G}$ is determined within our Solar System, due to dark and baryonic matter of the Milky Way, MW.\footnote{By Solar System we mean the Sun together with all the planets,  asteroids, Kuiper belt objects, etc.} 
Its magnitude relative to the centre of the Galaxy, CG, at the distance of the Sun is
\begin{equation}
F_{G} = 1.8 \times 10^{-10} m/s^2
\label{eq:HForcevalue}
\end{equation}
per unit mass.  In a Sun-centered coordinate system, $F_G$ is reduced by a factor of approximately $\mathcal{O}(10^{-7})$ since it is measured as a tidal force. 
$F_G$ is dominant over the gravitational force of the Sun on a body moving sufficiently far from the Sun, say 10s of thousands of AU. If $P$ is moving in interstellar space near the Solar System at distances sufficiently far from the Sun, the gravitational forces acting on $P$ are $F_G$ and $F_S$.

The derivation of (\ref{eq:HForcevalue}) is briefly summarized. The gravitational force acting on $P$, located near the Sun due to both dark and baryonic matter in MW can be modeled using a Hernquist model, applied in two steps.  
  
  The virial mass of MW, $M_{vir}$ as estimated in \cite{Watkins:2019}, 
 \begin{equation}
M_{vir} = 1.54_{-.44}^{+.75} \times   10^{12} M_{\odot},
\label{eq:GalaxyMass}
\end{equation}
for a virial radius $R_{vir} = 300$ kpc  (see also \cite{Callingham:2019}).   The Hernquist model gives the mass density $\nu(r) =(M_{vir}/2\pi )[d/(r(r+d)^3]$  at a distance $r$ from the Galactic Center, GC (see \cite{Hernquist:1990}).  The scale length $d=21.5$ kpc is assumed.  The virial mass includes both baryonic and dark matter without regard to the relative geometry of baryonic matter in the disc, bulge and stellar halo, and dark matter in the dark matter halo.

 The mass distribution is $M(r) = M_{vir}  r^2/(r+d)^2$ for $r \leq R_{vir}$.  $M_{vir} \equiv M(R_{vir})$.   This yields the total mass of MW for $r \leq r_s$, $M(r_s) = .077M_{vir}$,  where $r_s = 8.29$ kpc is the distance of the Sun to GC.
Thus, the mass within the sphere of radius 300 kpc for the galactic halo  of MW is used to estimate the mass within the smaller sphere of radius $r_s$.

Next, the gravitational force for $r \leq r_s$ is computed.  This uses a Hernquist model defined for the less massive stellar component of MW.  The force at a point ${\bf x}$ within this spherical region is
\begin{equation}
{\bf{F}_{G} }({\bf{x}})=  - \frac{  GM(r_s){\bf{x} }} {r(r+ b)^2},
\label{eq:GalForce}
\end{equation}
where ${\bf{x}} = (x_1, x_2, x_3)$ is relative to $GC$ at the origin,  $r = |{\bf{x}}| = (x_1^2 +x_2^2 +x_3^2)^{1/2} \leq r_s$, and where $b$ is a new scale length for the stellar halo of MW, $\leq r_s$.  $G$ is the Newtonian gravitational constant. $b=1$ kpc.  $F_{G} (r) = GM(r_s) (r+b)^{-2}$ is the magnitude of $(\ref{eq:GalForce})$. 
Setting $r=r_s$ yields (\ref{eq:HForcevalue}).

 It is noted that in $BG22$, $F_G(r_s)$ is also estimated using both the Navarro-Frenk-White (NFW) model (see \cite{Navarro:1996}) for dark matter distribution in MW together with a point-mass Newtonian force model for the baryonic contribution, whose value agrees with (\ref{eq:HForcevalue}). The multi-step Hernquist model yields a formulation more suitable for the purposes of this paper.

This model accurately estimates the velocity of the Sun about GC. 
The circular velocity of the Sun about GC  is defined to be $V_{C,s} = \sqrt{GM(r_s)/r_s}$. This yields,
$V_{C.s} = 248$ \ km/s, lying within the acceptable range $220-250$ km/s (see \cite{Watkins:2019}). \\

\noindent
{\it Point-Mass Newtonian Approximation} 
\medskip

$F_G$  can be approximated by a classical point-mass Newtonian force for MW, $P_{MW}$, with mass $M(r_s)$ placed at CG. This plays a key role in a simplified modeling near our Solar System for the dynamics of motion for a test particle, $P$, due to the 
gravity of MW and the Sun.  It is assumed that $P$ will move within a few 10s of LY from $S$.\\

\noindent Approximation of $F_G(r)$ \  \ For $r$ near $r_s$, the gravitational force due to MW is approximated as
\begin{equation}
F_G (r)  \approx  \frac{G M(r_s)}{r^2}.
\label{eq:ForceNewtonianApprox}
\end{equation} 
This is a point mass Newtonian approximation for $F_G(r)$, where MW is represented as point mass $P_{MW}$ at CG.\\

 $r$  {\it near} $r_s$ means that $r$ is within $4$ LY to $r_s$, or equivalently, that $P$ is within approximately $4$ LY so as not to extend to the distance of the closest star, Proxima Centauri at $4.2$ LY.
  In the units of kpc which are used for the length scale in $F_G(r)$, where a distance of $1$ corresponds to $1$ kpc, then $4$ LY represents approximately $.0001$.
\medskip
 (\ref{eq:ForceNewtonianApprox}) is verified by noting that 
\begin{equation}
(r+b)^{-2}  = r^{-2 }(1 + x)^{-2},
\label{eq:Equation}
\end{equation}
where $x = b/r \approx b/r_s = 1/8.29  \approx .12 < 1$, and $x$ is dimensionaless. Thus, $(1+x)^{-2}$ can be expanded in a convergent binomial series,
$(1+x)^{-2} = 1  - 2x +  \mathcal{O}(x^2)$  .
This implies, 
\begin{equation}
F_{G}(r)  =  \frac{G M(r_s)}{r^2}(1 - 2x + \mathcal{O}(x^2)).
\end{equation}
The leading term  yields the value given by (\ref{eq:HForcevalue}). The second term has fractor $2x  \approx .24$. This has the effect of reducing (\ref{eq:HForcevalue}), and
hence  (\ref{eq:ForceNewtonianApprox})  for $r$ near $r_s$ by an approximate order of magnitude, which given the small size of (\ref{eq:HForcevalue}) can be omitted as long as $r(t) \approx r_s$.
The higher order terms in the binomial
expansion reduce (\ref{eq:HForcevalue}) to even higher orders of magnitude.   This yields ({\ref{eq:ForceNewtonianApprox}).  

It is noted that ignoring the terms $-2x +\mathcal{O}(x^2)$ can be done only because (\ref{eq:HForcevalue}) is already small. Dropping these terms makes sense for sufficiently short time spans where their effect
on the dynamics of the motion of $P$ is negligible.  

\section{Modeling: Restricted Three-Body Problem} 
\label{sec:RestrictedProblem} 

 \noindent A simplified model is made for the dynamics of a test particle, $P$, as it moves near the Sun due to the gravitational forces of both MW and the Sun. $P$ could be a comet, rougue planet, for example.
  
It is assumed that $S$ movies about $P_{MW}$ at GC  in a circular orbit with constant radius $r_s$, constant circular velocity $V_{C,s}$, on a plane $\Sigma$ that passing through$P_{MW}$ and that $P$ is contrained to move on $\Sigma$.\footnote{It is noted that both $P_{MW}$, $S$ perform mutual circular motion about their common center of mass, essentially at CG.} These assumptions make sense relative to the galactic scale, locally for the motion of $P$ near $S$ (assumed within 10 LY).  It is also assumed that the only forces acting on $P$ are the gravitational forces due to  point masses $S$, $P_{MW}$ and the centripetal force due to the rotation of $S$ about $CG$ . 

Inertial coordinates on the plane $\Sigma$ are ${\bf{x}} = (x_1, x_2)$, centered on $P_{MW}$.   The mass of $P$ mass is negligible with respect to the Sun, taken to be 0. The mass of $P_{MW}$ is $M(r_s) = 1.19 \times 10^{11} M_{\odot}$. The mass of $S$, $M_S = 1 M_{\odot}$.  

 Thus, $P$ moves on the Sun's plane of motion about the center of the Galaxy. In the real setting, the Sun doesn't move on a plane, but rather oscillates above and below the galactic equator. The velocity of its motion, $248$ km/s, described in Section 2, is a good approximation based on observational data,

 These assumptions define the classical planar circular restricted three-body problem for the motion of $P$, going back to H.  Poincar\'{e} (\cite{Poincare:1899}).   It is convenient to choose a rotating coordinate system ${\bf y} = (y_1, y_2)$ that rotates about $P_{MW}$ with angular velocity $\omega = V_{C,s}$ in the $\Sigma$-plane.  In this system, $S$ is fixed and lies on the $y_1$-axis.  The $y_1$-axis passes through both $S$ and $P_{MW}$.
 
 The following nomalizations can be made without loss of generality: \  $\omega = 1$, \ mass of $S =  \mu$, where $\mu =  M_{\odot}/( M(r_s)  +  M_{\odot}) \approx  8.4 \times 10^{-12}$, $G=1$, the distance from $P_{MW}$ to $S$ is $1$, where $S$ is located on the $y_1$-axis at $y_1 =1$ and $P_{MW}$ is at the origin.
The center of mass between $P_{MW}$ and $S$ is at $(\mu, 0)$.  Under this scaling, the distance of the center of mass to $P_{MW}$ is just a few thousand   km.
 
    The differential equations for the motion of $P$ are given in the more general case for any two bodies $P_1, P_2$, where $P_1$ is the more massive primary and $P_2$ is less massive secondary (\cite{SiegelMoser:1971}).   The motion of $P$ for ${\bf y}(t)$ as a function of time, $t$,  is given by the system of differential equations,
   \begin{equation}
   \ddot{y}_1 - 2 \dot{y}_2  = \Omega_{y_1} ,  \hspace{.1in}
   \ddot{y}_2 +2 \dot{y}_1  = \Omega_{y_2} ,   \hspace{.1in},
       \label{eq:DEs}
   \end{equation}
  where $\Omega = (1/2) [(y_1-\mu)^2  + y_2^2]  + (1-\mu){r^{-1}_1} + \mu{r^{-1}_2} + (1/2)\mu(1-\mu)$, 
    and $r_1^2 = y_1^2 +y_2^2 $, $ r_2 ^2 = (y_1 - 1)^2 + y_2^2 $, $\Omega_x\equiv \partial \Omega / \partial x$,  $^. \equiv$ d/dt.  
  
  The Jacobi energy integral is  
  \begin{equation}
  J = J({\bm{y, \dot{y}}}) = 2 \Omega - |{\bm{\dot{y}}}|^2,  
  \label{eq:JacobiEnergy}
  \end{equation}
  $|{\bm{\dot{y}}}|^2 =  (\dot{y}_1^2 + \dot{y}_2^2 )$ .  Along a trajectory, ${\bm{\psi}} (t) = ({\bm{y}}(t), {\bm{\dot{y}}}(t) )$, $t \in \mathbb{R}$,\ $J({\bm{\psi}} (t)) = C$, the 
  Jacobi constant. The set $\{({\bm{y, \dot y}}) | J=C\}$ defines a three-dimensional Jacobi surface.  
  \medskip

      
The motion of $P$ is dynamically understood about $P_1$, starting with elliptical initial conditions for $\mu$ small, by the Kolmogorov-Arnold-Moser (KAM) Theorem (\cite{SiegelMoser:1971}), provided the ratio of the two frequencies, $\omega$ and $\omega^*$ (the frequency of $P$ for $\mu=0$) are sufficiently irrational. 
Then the trajectory of motion for $P$, ${\bm{\psi}} (t)$, $t \in \mathbb{R}$,   in four-dimensional position-velocity space, $({\bm{y, \dot{y} } })$, ${\bf {y}} = (y_1, y_2), {\bm  \dot{y}} = (\dot{y}_1, \dot{y}_2)$,  on fixed three-dimensional Jacobi energy surfaces lie on two-dimensional invariant tori for $\mu > 0$ sufficiently small. It is quaisperiodic. The motion between the tori is chaotic, but also stable since the tori separate the phase space on each Jacobi energy surface.  

However, for motion about $P_2$, the general motion for $P$ starting with Keplerian elliptical initial conditions is not dynamically understood, unless the motion is arbitrarily  close to $P_2$ for $C$ sufficiently large where the KAM theorem is valid (\cite{Kummer:1979}). This situation isn't realistic and not in the range of $C$ required in this study. 

New results in $B24$ on the weak stability boundary and its analogy with the boundary of a Mandelbrot set give an understanding of some of the dynamics about $P_2$, including permanent capture $P$ and the nonexistence of KAM tori for a suitable range of $C$.\\

   \subsection{Assumptions, Solar System Extent, Dark Matter, Lagrange Points}
   \label{subsec:InitialAssumptions}
\medskip

The value of $C$ determines the geometry of the Hill's regions in $y_1, y_2$ coordinates were $P$ is able to move \cite{Szebehely:1967}, \cite{Conley:1968}.  They are defined by  ${H}(C) = \{ {\bf y} | 2\Omega \geq C \}$. Let $C_i, \  i=1,2,3,4,5$ be the values of $C$ corresponding to the Lagrange-Euler points $L_i$, respectively, $3 =C_5 = C_4 < C_3 < C_2 < C_1$. For $C \geq C_1$ $P$ cannot pass between $P_2$ to $P_1$ since the Hill's regions about $P_1$ and $P_2$, $H_1, H_2$, repectively, are not connected. For $C$ slightly less than, but not equal to $C_1$, $C \lessapprox C_1$, the Hill's regions about $P_1, P_2$ are connected near $L_1$ in a channel $\mathcal{C}_1$ about a Lyapunov periodic orbit, $\gamma_1$. When $C$ decreases to $C \lessapprox C_2$ there is another channel $\mathcal{C}_2$ about $L_2$ connecting $H_2$ to the outer Hill's region, $H_O$ about both $P_1, P_2$. Another Lyapunov orbit, $\gamma_2$, appears in $\mathcal{C}_2$.

The range of $C$ is assumed to be $C_a < C < C_1$, where $C_a < C_2$ is defined in Section \ref{sec:PermCap}.  Thus, there are channels  $\mathcal{C}_i, 1=1,2$ on both sides of $P_2$ near $L_1, L_2$, respectively,  that $P$ can pass through.    

It is assumed $P_1 = P_{MW}, \ P_2 =S$, and  that $P$ moves from $H_1$ or $H_O$, defined to be interstellar space, into $H_2$ about $S$. $H_2$ defines the extent of the Solar System for this study.  This implies that the distance of $L_1, L_2$ to $S$, $\rho_1, \rho_2$, respectively,  approximates the extent of the solar system. For the small value of $\mu$, $\rho_1, \rho_2$ are approximately the same.  
\medskip 
 
  The distance of $L_1, L_2$ to $S$ can be estimated.   If this distance is labeled $\rho_1, \rho_2$, respectively,  then for any $P_1, P_2$ with $\mu$ sufficiently small,  
   \begin{equation}
  \rho_1 \approx \gamma (1- (1/3)\gamma)  , \hspace{.1in}  \rho_2 \approx \gamma (1+ (1/3)\gamma) ,
  \label{eq:LagrangeDistance}
  \end{equation}
\medskip
 within $\mathcal{O}(\gamma^3)$, where $\gamma = (\mu/3)^{1/3}$ (\cite{Belbruno:2004}, page 130)  (see also \cite{Szebehely:1967}). $\mathcal{O}(\gamma^3)$  is negligibly small given the magnitude of $\mu$.
\medskip
From the value of $\mu$,  $\rho_i$ can be computed, where $\rho_i \approx  \mathcal{O}(10^{-4})$, and converting to LY, 
\begin{equation}
\rho_1 \approx   3.81088        \hspace{.1in}   \rho_2 \approx  3.81092,
\label{eq:rho-real}
\end{equation}
$\rho_1 \lessapprox \rho_2$.  These distances correspond to approximately $241,005.5$ AU.  In scaled units, the distance $d$ from $S$ to either $L_1$ or $L_2$ is given by $d  \approx 1.41 \times 10^{-4}$.   
\medskip\medskip

Thus, the Solar System extends to approximately $3.81$ LY from $S$.   \\

It is noted that if dark matter were not included for $M_{vir}$ then the value of $3.81$ LY would be substantially larger: \  \ The baryonic mass of the Galaxy is estimated as 
$M_B = 5 \times 10^{10}M_{\odot}$ (\cite{Licquia:2013}). Using this for $M{vir}$, then $M(r_s) \equiv M(r_s, B) =  M_B  r_s^2/(r_s+d)^2 = .077 M_B$ and $\mu \equiv  \mu_B =  M_{\odot}/( M(r_s)  +  M_{\odot}) \approx 2.60  \times 10^{-10}$. This yields, $\rho_i \equiv \rho_{i,B} \approx  11.97$ LY. 
\\

{\it Conclusion  \hspace{.1in} If we model the galaxy with enough mass to get the correct circular velocity, and if we use the point mass approximation, then the extent of
the Solar System is 3.81 LY. If we only consider part of the galaxy by not including dark matter and hence consider an unrealistically lower mass, then we would get a larger
extent. }\\

The openings about $\gamma_i, \ i=1,2$ connecting $H_2$ with $H_1, H_O$, respectively, are small channels that $P$ can move through to pass from $H_1$ or $H_O$ into $H_2$.  They are labeled $\mathcal{C}_i, \ i=1,2$, respectively. 
\\

It is assumed that the motion of $P$ starts in interstellar space in $H_1$ or $H_O$ outside $\mathcal{C}_1$ or $\mathcal{C}_2$, respectively.  The initial distance of $P$ from $S$ is assumed to be approximately $4$ LY.  This distance is chosen to be less than the distance to Proxima Centauri.

\section{Permanent Weak Capture Mechanism} 
\label{sec:PermCap}

A $P_2$-centered coordinate system ${\bf Y} = (Y_1, Y_2)$ is adopted, by the translation $Y_1 = y_1 -1, Y_2=y_2$. 
The two-body Kepler energy of $P$ with respect to $P_2$,  $E_2$, is given in Section \ref{subsec:KeplerEnergy}.  Although this is stated
for general $P_1, P_2$, we are assuming $P_1=P_{MW}, P_2=S$.  

Let ${\bm{\psi}} (t) = ({\bm{Y}}(t), {\bm{\dot{Y}}}(t) )$ be a solution to the restricted problem for $P$ defining a trajectory as a function of time, $t$, in the $P_2$-centered rotating system  ${\bf{Y}} = (Y_1, Y_2)$.   The Kepler energy along the trajectory is defined as $E_2(t) = E_2({\bm{\psi}} (t) )$. Let $\tilde{J}(\bm{Y}, \bm{\dot{Y}})$ be the Jacobi integral in these coordinates.

 Weak capture is defined along ${\bm{\psi}} (t)$ at a time $t=t_1$: \
If $E_2(t_0) \geq 0$  and at a later time, $t= t_1 > t_0$,  $E_2(t_1) <  0$, then $P$ is weakly captured by $P_2$ at $t=t_1$.  Weak capture is used, for example, in  \cite{Napier2:2021} (see page 3, defining capture when the Kepler energy goes below zero).
\medskip

$n$-stable cycling motion is defined assuming $P$ starts on a reference line $L(\theta)$  from $P_2$, at $t=t_0$, making an angle $\theta \in [0, 2\pi]$ with the $Y_1$-axis. The initial conditions of ${\bm{\psi}} (t_0)$ on $L(\theta)$ with resepct to $P_2$ it is assumed:\  $E_2(t_0) < 0$, the eccentricity $e \in [0,1)$ for $P$ and the initial velocity vector is perpendicular to $L(\theta)$, assuming counterclockwise motion. $P$ cycles about $P_2$ $n$ times, $n \geq 1$, without cycling around $P_1$. Each of the $n$ crossings with $L(\theta)$ is transversal with $E_2(t) < 0$.   Otherwise it is $n$-unstable \ (see $GG07$, \cite{TopputoBelbruno:2009}, \cite{BelbrunoGideaTopputo:2010}($BGT10$ for brevity) ). $n$-unstable cycling can happen several different ways: \ (i.) $P$ cycles about $P_1$ before making $n$ cycles about $P_2$, \ (ii.) A crossing with $L(\theta)$ is tangential, \ (iii.) $E_2 \geq 0$ at a crossing, \ (iv.) \ $P$ does not return to $L(\theta)$ prior to $n$ cycles.

The points along $L(\theta)$ that are initial conditions for trajectories at the boundary between $n$-stable  and $n$-unstable motion about $P_2$ in $Y_1, Y_1$-space are parameterized by $e,\ \theta, \ n$. It is labeled $W_n(\theta, e)$ and called the {\it $n$th weak stability boundary}. Taking the union over all $e\in [0, 1)$ yields $W_n(\theta)$, of particular interest for this paper. The Jacobi integral value $C$ varies over the points of $W_n(\theta,e)$. The points of $W_n(\theta,e)$ are themeselves $n$-unstable. 
(The case of $n=1$ was first defined in \cite{Belbruno:2004})

It is interesting to relate the $n$-cycling motions with realistic numbers. The distance of 3.81 LY from $S$ is of particular interest for this paper. This is the distance to $L_1, L_2$ for $C=C_1,C_2$, respectively. For $C \lessapprox C_1$, the channel $\mathcal{C}_1$ appears about $L_1$.  For $C \lessapprox C_2$ a channel  $\mathcal{C}_2$ appears about $L_2$ in addition to the one about $L_1$. 

The interior of the Hill’s region $H_2$ about $S$ extends out to about $3.8$ LY for $C \lessapprox C_{1,2}$.  When $C \lessapprox C_2$,  $P$ can be captured from interstellar space ($H_1$, $H_O$) into our Solar System $H_2$.    As is shown in $B24$, the dynamics of motion is sensitive at a distance near the Hill’s boundary, at about $3.8$ LY,  for the energy range $C \lessapprox C_2$ extending to $C \lessapprox C_1$ due to the intersections of manifolds extending from the Lyapunov orbits in the channels.   
When performing cycling motion from $L(\theta)$ near this distance, the cycles take a long time to move about $S$ if one starts, for example, with an osculating circular state. If $P$ could perform one complete near circular cycle with near circular velocity $61$ m/s, without going unstable due to the gravitational perturbation due to $P_{MW}$, the time of a cycle would be $117$ Myr.  In a purely  mathematical context, one could talk about infinitely many cycles, but in realistic terms, $10$ cycles is already about $1$ billion years. If $P$ were mathematically determined to be on a permanent weak capture trajectory, taking infinitely many cycles to approach a weak capture point about $S$, in practical terms, one would only need at least a few cycles.  It is noted that starting the circular orbit say $100,000$ AU from $S$ has a cycling time of $32$ Myr.


In $BGT10$,  it is shown through numerical and analytic analysis (semi-analytic) that $W_n(\theta)$ is equivalent to the points on the stable manifolds associated to $\gamma_i$, $i=1,2$ in the Hill's region $H_2$ about $P_2$ satisfying $\dot{r} = 0, E_2(\bm{Y}, \bm{\dot{Y}}) < 0$, where $r = |\bf{Y}|$.  These manifolds are labeled $W^s(\gamma_i)$. Set $\Lambda = \{ {\bm{Y}}, {\bm{\dot{Y}}} \ | \ \dot{r} = 0, E_2 <0 \}$.  Thus, the points of $W_n(\theta)$ belong to $\Lambda$. The points of $W_n(\theta)$ are obtained as the intersections of $W^s(\gamma_i)$ on a two-dimensional surface of section  $S_{\theta}$, with coordinates $r, \dot{r}$, for each fixed $\theta$ and a fixed admissible energy value within $C \in  \{C_a < C < C_1\}$.  $C_a$ is the largest value of $C$ where $W^s(\gamma_i$) intersects with $P_2$.

The appearance of $W_n(e)$ (= union of $W_n(\theta, e)$ over $\theta \in[0, 2\pi]$) was noted in $GG07$, $BGT10$ to appear fractal for a small set of $e$, but wasn't justified. 

The fractal structure is proven in $B24$ for the limiting set $W'(\theta) = \lim_{n \rightarrow \infty} W_n(\theta)$ over infinitely many cycles about $P_2$. 
More precisely, it is shown that $W'(\theta)$ on $S_{\theta}$ is a Cantor set on each Jacobi energy surface $\{\tilde{J}(\bm{Y}, \bm{\dot{Y}}) =C\}$ for each $C$ in a special subset of $C_a < C < C_1$. Besides the Cantor set being everywhere disconnected, it is also self-similar, so that when
any part of  $W'(\theta)$, no matter how small, is magnified, it equals all of $W'(\theta)$. Each of the points of $W'(\theta)$ have one dimensional stable and unstable 
manifolds. Let $W'$ be the union of $W'(\theta)$ over all $\theta \in [0,2\pi]$.  The points of $W'$ belong to $\Lambda$. Trajectories ooving within $W'$ are chaotic, and trajectories moving near $W'$ are generally unstable.  
\medskip

\noindent
{\it Comparison to Other Approaches}
\medskip

It is noted that in \cite{Napier1:2021} permanent capture is considered to be capture for $1$ billion years.  This is a different notion that what we are considering in this paper, where capture is for infinite time.  The capture considered in this paper is permanent over infinitely many cycles as time $t \rightarrow \infty$ that converges to a point of a special fractal set. This is a different mechanism than what is considered in \cite{Napier1:2021}. In that paper, many trajectories are propogated using a Monte Carlo method that does not study the dynamical nature of permanent weak capture. This paper is concerned with understanding the dynamics associated with permanent weak capture, where it exists, and its theoretical properties.   It is also worth noting that \cite{Napier1:2021, Napier2:2021, Hands:2020} are not modeling the gravitational perturbation of the Galaxy but rather Jupiter, ($S$, Jupiter, $P$). This is a different model used here, $S$, Galaxy, $P$.  

It is noted that at the distance of $3.8$ LY from $S$,  the magnitudes of the gravitational force on $P$ due to $P_{MW}$ and Jupiter are of the same order, being $1.44 \times 10^{-17}$ m/$s^2$ and $9.8 \times 10^{-17}$ m/$s^2$,  respectively.  This means that if $P$ is moving slowly at say $100$ m/s  relative to $S$  at this distance, the $P_{MW}$ gravitational force competes with Jupiter’s gravitational force. In our model, the Sun's gravity is used to model the gravitational force of the Solar System since $P$ is moving $3.8$ LY from the $S$. The gravitational perturbation due to $P_{MW}$ is the main perturbation at that distance.  The Galaxy perturbation is less important when $P$ moves much closer to $S$ or with higher velocities.  
\medskip

\noindent
{\it Permanent Capture Set}
\medskip

A trajectory ${\bm{\psi}} (t)$ starting on $W'$, starts with $E_2 <0$  and generally on this trajectory $E_2(t) \geq 0$ for some $t$ since $W'$ consists of unstable cycling points. This occurs for the unstable cycling points satisfying $n$-unstable conditions (i.), (iii.) as $n \rightarrow \infty$ which we restrict to.  By the symmetry condition, $(Y_1(t), Y_2(t), \dot{Y}_1(t), \dot{Y}_2(t)) \rightarrow (Y_1(-t), Y_2(-t), \dot{Y}_1(-t), \dot{Y}_2(-t)) $ this yields a trajectory, ${\bm{\bar{\psi}}} (t)$ that starts with $E_2 \geq 0$ and movies to weak capture at a point $p$ on $W'$.  This occurs with ${\bm{\bar{\psi}}} (t)$ asymptotically approaching $p$ as $t \rightarrow \infty$ by cycling about $P_2$ infinitely many times. Trajectories can be chosen to satisfy (i), (iii) due to the chaotic structure of $W'(\theta)$ on $S_{\theta}$ giving infinitely many possible trajectory itineraries.

However, the points of $W'$ satisfying (i.) are of particular interest since in this case  the symmetric trajectory ${\bm{\bar{\psi}}} (t)$ enters the Hill's region $H_2$ through channels $\gamma_i$, $i=1$ or $2$ at $180$ degrees on either side of $P_2$ and goes to weak capture to a point of $W'$. 

Let $\bar{W}'$ be the subset of points of $W'$ containing only those points satisfying condition (i.).  Therefore, this comprises the subset of permanent capture points of $W'$.  Trajectories going to these points start from outside the Hill's region $H_2$,  enter $H_2$ through channels $\mathcal{C}_i$ and go to permanent capture at points on the set $\bar{W}'$ after infinitely many cycles about $P_2$. 
\medskip

\noindent {\it Applications to $S, P_{MW}$ }
\\

Choosing $P_1 = P_{MW}, \ P_2 = S$, then the set $\bar{W}'$ about $S$ respresents points that are permanently captured from interstellar space.  In this case, a body $P$  starts its motion in $H_1$ with a positive Kepler energy with respect to $S$.  We assume it starts near the channel opening $\mathcal{C}_1$ in the $H_1$ region, defined to be in interstellar space, where the Jacobi energy is assumed to be $C \lessapprox C_1$ or $\lessapprox C_2$. This restricts the magnitude of velocities $P$ can have relative to the Sun. The velocity will be small. We are assuming on the order of $10$s of meters per second. If $C$ is very close to $C_1$ and $P$ goes through the channel, it exits into the $H_2$ region at about $3.8$ LY from the Sun and starts to move about $S$. If its velocity is $61$ m/s it will start to move in an osculating circular orbit. As noted previously it will take 117 Myr to circle $S$. However, since the magnitude of the gravitational forces due to the Galaxy is $\mathcal{O}(10^{-17})$,  the trajectory for $P$ will start approximately circular and then over time it will deviate from the circular state. It will spiral about $S$ and in the idealized mathematical problem, over infinitely many spirals it will asymptotically approach a point far from the Sun, where $\dot{r}=0, E_2 <0$.    This point lies in $\bar{W}'$ and is a point in a Cantor set, dynamically a hyperbolic point, with stable and unstable manifolds on a two-dimensional surface of section for a given value of the polar angle $\theta$ from the $Y_1$-axis. This means that the motion of $P$ is sensitive as it approaches this point.  In fact,  it asymptotically approaches this weak capture point as time $t \rightarrow \infty$, performing infinitely many cycles about $S$. It will pass close to infinitely many such Cantor points for different values of $\theta$.  
As described in $B24$, for each $\theta$, there is a Cantor set of infinitely many possible weak capture points with $\dot{r}=0, E_2 <0$. Thus there are infinitely many such Cantor sets of infinitely many possible weak capture points for $0 \leq \theta \leq 2\pi$.  However, a trajectory will converge to just one of them. But to get there, it has to move through regions containing infinitely many Cantor points with stable and unstable manifolds making the motion very sensitive.    
  It is shown in $B24$ that the colllection of infintely many Cantor sets of possible capture points for $0 \leq \theta \leq 2\pi$ comprises the weak stability boundary for infinitely many cycles. It bounds stable cycling motion about $S$ for infinitely many cycles for trajectories starting at distances from $S$ less than the boundary location. And for trajectories starting outside this boundary, they will generally be ejected from the Solar System.

The distribution of the points of $\bar{W}'$ about $P_2$ in general is not known, and represents an interesting problem. . This represents a substantial numerical study beyond the scope of this note. Computing $W_n$ even for smaller values of $n =8$ (the highest computed) takes a significant effort. The Cantor points of $W'$ for high values of $n$ would be difficult to discern.  The points of $\bar{W}'$ would be among the points of $W'$, and even harder to discern. However, as is shown in $B24$,  the general structure can be theoretically determined. As seen in the following text, even though the distribution is not known, the probability of capture can be estimated.

It is shown in $B24$ that $W'$ projected into $(Y_1,Y_2)$-space is the boundary of a region, $\hat{S}$, about $P_2$ where the motion cycles about $P_2$ infinitely many times in a stable manner.  That is, it consists of points as limits of $S_n$ as $n \rightarrow \infty$, labeled $\hat{S}$.  On $S_{\theta}$, the points of $W'(\theta)$ are themselves unstable points for infinitely many cycles and lie at the boundary between stable and unstable cycling motion, whereas the points of $\hat{S}(\theta)$ are stable for infinitely many cycles about $P_2$. The points of $\hat{S}$ also belong to $\Lambda$.
This shows that $W'$, and hence $\hat{W}'$, represent boundary points for stable cycling motion about $S$. $W'$ is where stable cycling motion can no longer occur.\\

It is noted that if a trajectory ${\bm{\psi}}(t)$ starts on a point of $W'(\theta)$ for $t=0$,  then for $t >0$, ${\bm{\psi}} (t)$ cycles about $P_2$ in $(Y_1, Y_2)$-space remaining on $W'$. It repeatably passes though points of $W'(\theta)$ on $S_{\theta}$. This yields a map $\Phi$ on $S_{\theta}$, mapping $W'(\theta)$ into itself.
Interestingly, $\hat{S}$ is shown to share similar properties to a Mandelbrot set for the much simpler complex map $Q_c(z) = z \rightarrow z^2 +c$ starting at $z=0$ in the complex plane, where $z, c$ are both complex numbers.  The Mandelbrot set, $M$, consists of those points in the complex $c$-plane that yield bounded iterates of $Q_c(0)$ over infinitely many iterates. The boundary of the Mandelbrot set is a connnected fractal curve whereas the boundary of $\hat{S}$, which is $W'$, is totally disconnected for each $\theta$.  
\\

\noindent{\it Probability of Permanent Weak Capture, Capture Cross Section}
\\

 The probability of permanent weak capture and the properties of the permanent weak capture cross section can be estimated in the case of a rogue planet or any other object such as an asteriod or  comet.
It turns out, as described in the following, the capture cross section can be estimated for three-dimensional motion, and that the width of the channels $\mathcal{C}_i, i=1,2,$ for planar motion considered thus far gives an estimate of the size of the capture cross section for three-dimensional motion. Consideration of the dynamics of motion near the Lyapunov orbits  $\mathcal{\gamma_i}, i=1,2,$ together with the geometry of the cross section yields the probability of weak permanent capture, assuming the Jacobi constant $C$ is suitably restricted to the set $A =\{C \ | \ C \lessapprox C_2\}$. However, it also turns out that not every point of this cross section yields a weak permanent capture trajectory.  The set of points that yield weak permanent capture trajectories for a given value of $C$ is a set of Cantor points comprising a fractal set, where the velocity magnitude for $P$ varies across these points and across the set $A$. }

The case of two-dimensional motion is considered first. As described preiously, permanent weak capture trajectories exist for $C_a  <  C  < C_1$. However, as $C$ decreases from $C_2$ towards $C_a$, the width of $\mathcal{C}_i$ increases significantly. To control the width of the channels connected to the $\gamma_i$, it is noted first that $C_1$ and $C_2$ agree to nine digits. Thus, $C_2 \lessapprox C_1$.  If $C$ is restricted to the set $A$, then the widths of $C_1, C_2$ are approximately the same. This means that it is sufficient to just consider the case where $P$ moves from $H_1$ into $H_2$ through $\mathcal{C}_1$. (The estimates for the case where $P$ moves from $O$ through $\mathcal{C}_2$ into $H_2$ will be approximately the same.)  It is noted that $\mathcal{C}_1$ is a two-dimensional set in the $Y_1, Y_2$-plane symmetric about the $Y_1$-axis, centered about $\gamma_1$, which moves in a retrograde fashion about the $L_1$ location, symmetric about the $Y_1$-axis and extending from the lower boundary of $\mathcal{C}_1$ to the upper boundary. 

The width of $\mathcal{C}_1$ for a given $C \in A$ is defined to be the vertical width of the Lyapunov orbit about the $L_1$ position, described below. For each $C \in A$, there is a Cantor set of infinitely many hyperbolic points on the two-dimensional Poincar\'e section $S_{\theta}$, in the four-dimensional phase space,  for a value of $\theta$ corresponding to the $L_1$ position as measured from $P_2$.  $S_{\theta}$ intersects $\mathcal{C}_1$ in a line, $\mathcal{\hat{L}}$, through the $L_1$ position, which is not vertical, when projected into physical space. Trajectories passing from $H_1$ to $H_2$ will pass through $\mathcal{C}_1$ and intersect the line segment $\mathcal{\hat{L}} \cap {\mathcal{C}_1}$.   A subset $B$ of the Cantor points on $S_{\theta}$ for the given value of $C \in A$ belong to a subset of $W'(\theta)$,  which by symmetry with respect to the $Y_1$-axis, $Y_2 \rightarrow -Y_2$ and $\dot{Y}_1 \rightarrow -\dot{Y}_1$,  yield points where permanent weak capture trajectories exist in phase space from $H_1$ and intersect $S_{\theta}$. In physical $Y_1, Y_2$- space, these trajectores pass into $H_2$ from $H_1$, intersecting $\mathcal{\hat{L}} \cap {\mathcal{C}_1}$.

The line segment $\mathcal{\hat{L}} $ gives a line segment passing through the $L_1$ location that estimates a capture cross section in phase space. The points of this region in physical $Y_1, Y_2$-coordinates can be projected to the vertical position at the $L_1$ location, relative to $Y_1$-axis by just following the flow of trajectories. This makes a vertical line segment, $\mathcal{L}$,  passing through the vertical axis of the Lyapunov orbit and through the $L_1$ position, for the given value of $C$, inside $\mathcal{C}_1$ and extending to its boundary.  The length of $\mathcal{L}$  estimates the width of the channel.   

This segment, $\mathcal{L}$, estimates the length of a capture cross section $\mathcal{S}$ in the planar problem projected into two-dimensional physical space. But not every point on $\mathcal{L}$ corresponds to a point of a weak permanant capture  trajectory. It is only those points corresponding to the Cantor set for the given value of $C$ that that belong to $W'(\theta)$, comprising the set $B$. Thus, the projection of $\mathcal{S}$ onto $\mathcal{L}$ has a fractal structure with fractional Hausdorff dimension. As $C$ varies, and hence as the velocity of the trajectories vary, the set of points of $W'(\theta)$ on $\mathcal{S}$ for different values of $C$ will vary. 

The variation in the distribution of the weak permanant capture points on $\mathcal{L}$ varies with $C$.   The variation of the distribution of weak capture points on $\mathcal{L}$ as a function of $C$, or on $\mathcal{S}$, would need to be investigated numerically, which is beyond the scope of this paper. 

Each of the permanent weak capture points on $\mathcal{L}$ for trajectories will have different velocities of $P$ associated with them, that can be seen on $\mathcal{S}$. Not all the points of the two-dimensional set $\mathcal{S}$ belong to permanent weak capture trajectories for a given $C$. This implies not all the points of $\mathcal{L}$ correspond to permanent weak capture trajectories for a given $C$. If for a given $C$ a particular velocity magnitude is desired for a permanent weak capture transfer, say $v^*$, then one has to see if a point of  $\mathcal{S}$ exists with that velocity.  If not, then one would change the value of $C$ in the set $A$ and check again to see if $v^*$ can be achieved.

 In physical space for the planar problem in the $(Y_1, Y_2)$-coordinate system with $S$ at the origin, scaled to physical units,  the channel $\mathcal{C}_1$ opening in $H_1$ is approximately centered at the location $(-3.8, 0)$ (LY)  approximately where $L_1$ is located, which is replaced by a Lyapunov orbit \cite{Belbruno:2004, LlibreSimo:1985}. This orbit is retrograde, elliptical in shape,  centered at  $(-3.8, 0)$ and syymetric with respect to the $Y_1$-axis.   As it periodically cycles about $(-3.8, 0)$ it has a maximal position above $(-3.8, 0)$ at $(-3.8, b)$ and below at $(-3.8, -b)$, $b$ is the sem-major axis.  $w=2b$ is the width of the orbit and the width of the channel opening.  

The width can be estimated by determining $b$,  \cite{LlibreSimo:1985} (see page 123 for a formula for $b$ as a function of $\Delta C = C_1-C_2.)$.  $w$  is determined to be $7880$ AU. This is the length of the perpendicular vertical distance centered on the $Y_1$-axis going though $(-3.8, 0)$, with $3940$ AU above and below $(-3.8 ,0)$.  This is the length of the opening to the channel in $H_1$ that $P$ has to pass through for the  range of Jacobi energy considered, $C \lessapprox C_2$  This subtends an angle of $\delta \approx 1.87$ degrees when viewed from from S.  We assume P starts its motion $3.8$ LY to the left of $(-3.8,  0)$, near $ Q = (-7.6, 0)$ in the $H_1$ region. $P$ will also see the channel opening to the right at $(-3.8 ,0)$ as subtending the same angle $\delta$.    

 When $P$ starts it motion near or at $Q$,  locally its trajectory will be approximately linear since it is assumed to be sufficiently far from $S$. It will remain approximately linear up to the opening of $\mathcal{C}_1$. Thus, the probability $P$ will move in the desired direction to pass through the opening is $P_A = \delta/360 = .005$.    

$P_A$ is estimated for the planar problem. Its value is valid for the three-dimensional restricted three-body problem where motion out of the $(Y_1, Y_2)$-plane is possible, assuming the out of  $(Y_1, Y_2)$-plane distance of $P$ is sufficiently small. In the three-dimensional restricted problem, the Hill's regions are spherical in shape instead of circular, and the channels are cylindrical. The out of plane height of the channels is of the same order as the in plane width for the same range of Jacobi energy.

We assume that the trajectory for $P$ moves within the distance of $.05$ LY above and below the $(Y_1, Y_2)$-plane from $Q$ to the channel opening.  This constrains the motion of $P$ to a total out of plane distance of $.1$ LY. The angle $\delta$ is still approximately subtended  from above or below  $Q$ of the width of the channel opening.   The channel opening is now a rectangle, $\mathcal{R}$, perpendicular to the $(Y_1, Y_2)$-plane that has a distance of $.05$ LY above and below the plane, intersecting the plane in a line perpendicular to the $Y_1$ axis at $(-3.8, 0)$ of total width $7880$ AU.  
Under these assumptions the out of plane motion is small relative to $3.8$ LY  and  the approximate linear trajectory of $P$ after starting near $Q$ will enter the channel if it crosses $\mathcal{R}$. 

$\mathcal{R}$ serves as a capture cross section for those points on $\mathcal{R}$ yield a suitable velocty magnitude range determined by $C$. The area of $\mathcal{R}$ is $4.98 \times 10^7$ AU$^2$. In $B24$ it is shown that the set of weak capture points in general on sections in the phase space have zero topological dimension, but fractional Hausdorff dimension. This implies that there will be an infinite set of points on $\mathcal{R}$ that generate weak capture trajectories with a fractional Hausdorff dimension. This is not a smooth cross section where all the points can lead to permanent weak capture, but it is a large area. This is compared to the cross section obtained in \cite{Napier2:2021} described in Figure 2, page 4. This is obtained for a different model, $S$, Jupiter, $P$ (the galaxy gravitational force is not modeled) and for weak capture when the Kepler energy of $P$ necomes negative. It is not looking for permanent weak capture by infinite cycling, but rather a finite capture only.  The cross section is $\mathcal{O}(10^{7})$ AU$^2$ for velocities that are similar to the ones we consider, in the $10s$ of m/s. We obtain the same order, but in the text after this we obtain $\mathcal{O}(10^{8})$ with more realistic conditions. 

Next we estimate the probability a rogue planet will be within a $7.6$ LY cube centered at $S$ and show that it is $P_B = .21$. This is seen as follows: The average number of stars within a unit cubic volume around the Sun is estimated as $.004$ stars per unit LY \cite{Gregersen:2010}. On the other hand, it is estimated that there are $20$ times the number of rogue planets per star \cite{SumiKoshimoto:2023}. 
This gives $.6$ rogue planets in this $7.6$ LY cube. In order to have $1$ rogue planet within a cube, its dimension should be 12.7 LY about $S$. Within this larger cube we are interested in the probability $P$ will be within a cube of $7.6$ LY dimension about $S$ so it could be near $Q$.  Since the probability of being in a smaller cubic volume within the larger cube is a function of the relative volume,  the probability $P_B = (7.6)^3/(12.7)^3 = .21$.

The probability, $P_N$ of $P$ being in a small neighborhood of $Q$ so it still approximately sees the same subtended angle $\delta$ at $(-3.6, 0)$ is estimated. We assume a small cube centered at $Q$ with dimension of $.1$ LY.  One half of this cube is within the $7.6$ LY cube. This makes a region of dimensions, $.05$ LY  length (parallel to the $Y_1$-axis), width $.1$ LY (parallel to the $Y_2$-axis, and $.05$ LY above and below the $Y_1, Y_2$ plane. This gives a volume of $.05 \times .1 \times .1 = .0005$ LY$^3$.   This implies the probability of $P$ being within this small neighborhood is $P_N = .0005/(7.6)^3 = .0000011$.

The final probability to determine is the probability that $P$ will be in the required velocity magnitude range, $<v>$  so when it passes through $\mathcal{R}$ it will have a suitable velocity to be permanantly weakly captured.  This probabilty is denoted by $P(<v>)$.   

 In summary, an approximate probability that a rogue planet can be permanently weakly captured into our Solar System is given by
\begin{equation}
\mathcal{P} \approx P_A P_B P_N P(<v>) = 1.16 \times 10^{-9}  P(<v>)    .
\label{eq:prob}
\end{equation}
assuming it lies within a cube centered at $S$ of $12.7$ LY dimension,  and assuming it moves out of the  $(Y_1, Y_2)$-plane within $.1$ LY,  starting near $Q$.  $\mathcal{R}$ serves as a region in space to transversally cross to be weakly captured, for those points on it that yield the required velocity range for the range of $C$.  It serves as a capture cross section, of area $4.98 \times 10^7$ AU$^2$ for those points on it that correspond to the desired velocity values $<v>$ for $C \lessapprox C_2$. These points should have a fractional Hausdorff dimension.

$\mathcal{P}$ is conservative since the motion of $P$ is constrained within $.1$ LY out of plane. This constraint is fairly tight and a much larger out of plane motion could be assumed. This would require more careful analysis since $P$ could move to the boundary of $H_1$ before moving to $\mathcal{R}$, that is out of the scope of this paper. It's likely the value of $.1$ could be much larger, say $1$ LY, increasing $P_N = .5/(7.6)^{3} = 1.1 \times 10^{-3}$.  $\mathcal{P}$  is increased by three orders of magnitude to 
\begin{equation}
\mathcal{P} \approx  1.16  \times 10^{-6}  P(<v>)    .
\label{eq:prob2}
\end{equation}
This may be a more realistic estimate.  $\mathcal{R}$ has area $4.98 \times 10^8$ AU$^2$.  

 $P(<v>)$  is not currently known, but it is an observable. Searching for rogue planets with powerful telescopes, for example using the  JWST could help determine this.

Summarizing, the range of probabilities of weak permanent capture from (\ref{eq:prob}), (\ref{eq:prob2}) is given by,
\begin{equation}
\label{eq:probrange}
          1.16 \times 10^{-9}  P(<v>)  \  \lessapprox   \   \mathcal{P} \ \lessapprox  \  1.16  \times 10^{-6}  P(<v>)   .
\end{equation}

To determine the actual capture rate, the number of distinct rogue planets near our Solar System that enter a cube centered at $S$ of dimension $12.7$ LY should be estimated over a range of time, say $N$ rogue planets per year.  This is currently not known and is an observable. The JWST could be used to help determine this. Thus, the number, $n$ of rogue planets that could be permanently weakly captured over $M$ years would be  
\begin{equation}
n = N M \mathcal{P}.   
 \label{eq:numberrogue}
\end{equation}

\section{Capture of Interstellar Objects into the Solar System }
\label{sec:InterstellarObjects}

Objects in interstellar space outside $H_2$ beyond $3.81$ LY can be permanently captured about the Sun in the manner described in the previous section by passing through $\gamma_i, i=1,2$, and cycling to capture at a point of $\hat{W}'$ as $t \rightarrow \infty$. Such objects could be comets or rogue planets.

\subsection{Rogue Planets}
   \label{subsec:RoguePlanets}

All stars are born from collapsing interstellar clouds forming widely dispersed stars in open clusters or close stars in globular clusters \cite{LadaLada:2003}.  Our Sun was born in an initial open cluster of about 2,000 to 20,000 interacting stars before dispersing over time (\cite{ArakawaKokubo:2023}) and these stars must contain rocky planets, giant planets, and small bodies. 

The final architecture of  any solar system will be shaped by planet-planet scattering \cite{SumiKamiya:2011} in addition to the stellar flybys of the adjacent forming star systems since close encounters can pull planets and small bodies out of the system creating what are called rogue planets (\cite{BrownRein:2022}). It has also been predicted that our Sun captured comets from other stars in the birth cluster by this close encounter process (\cite{Levinson:2010}).  By the very nature of the creation of solar systems in clusters of stars many rogue planets must be created. 

In addition to rogue planets being generated during the formation process, planetary scientists have come to realize that our own Solar System has gone through the process of giant planet migration after an initial period (believed to be ~850 My) of stable planetary resonance has been broken. An early solar system containing five giant planets was proposed by \cite{Nesvorny:2011} after numerical models indicated that this is more likely to reproduce the architecture of the current solar system, reproducing the observed scattering of a number of Kuiper Belt objects, after ejecting a Neptune-sized planet. Currently, an observation campaign is underway to find the fifth giant planet, which is referred to as Planet 9 (\cite{Batygin:2016}), assuming the ejected planet is still bound to our Sun. The other possibility is that the ejected planet is a rogue. 

When taken together, planet ejection from early planet-planet scattering and stellar encounters and in the subsequent evolution of a multi-planet solar system should be common and supports the evidence for a very large number of rogue planets that are free floating in interstellar space that perhaps exceed the number of stars (\cite{SumiKamiya:2011}) but this assertion has been controversial. It is also noted that many planets in a solar system maybe liberated during a supernova explosion of the host star (\cite{Blaauw:1961}). There is very limited observational knowledge that leads to a better understanding of the number of rogue planets in our galaxy. Therefore, simulations of stars forming in clusters have been the primary tool used to obtain this number (e.g. \cite{Elteren:2019}, \cite{ParkerQuanz:2012}) which is still expected to be a significant component of the substellar population. With the advent of the James Webb Space Telescope (JWST) a robust census of rogue planets is possible since it is the first telescope with the capability to directly image rogue planets down to a Jupiter mass or even below.    
\medskip

 Since the passage of a rogue planet through our Solar System could cause gravitational disturbances to the planets and affect Earth's orbit about the Sun the capture cross section is calculated. The average number of stars within a unit volume around the Sun has been estimated as $0.004$ stars per cubic light year  \cite{Gregersen:2010}. This number varies with distance from the galactic equator and is higher as we move toward the galactic plane. Currently we are above the plane. The current estimate is that there are $20$ times the number of rogue planets (rp) per star \cite{SumiKamiya:2011} near the Sun producing about $0.08$ rp/LY$^3$. At the distance of $L_1$ this produces $0.3$ rp/LY$^3$ producing  capture cross sections of $4.98 \times 10^7$ AU$^2$, $4.98 \times 10^8$ AU$^2$ where there is a subset of points on these sections having suitable velocity values,  described in Section \ref{sec:PermCap}.

\subsection{Close Approach of Other Solar Systems}
   \label{subsec:Close Approach of Other Solar Systems}

Today the solar neighborhood, a volume with a radius of ~6 pc around the Sun, contains about 131 stars and brown dwarfs, several of which have known planets (\cite{Faria:2022}). All these stars move relative to the Sun and over time it is estimated that on average 2 stars per million years come within a few light years of the Earth, however, six stars are expected to closely pass by in the next 50,000 years (\cite{Matthews:1994}).  Since the outer boundary of the Oort Cloud is about 1.5 light years distance, stellar encounters must eject Oort Cloud comets inward toward, or away from, the Sun (\cite{Levinson:2010}). It is important to note that permanent capture, as described in this paper must also be occurring depending on the location of the passing star, along with it’s small body debris such as asteroids in addition to comets, as it passes by the Sun. It may turn out that, over time, the permanent capture of interstellar objects associated with passing stars near $H_2$ have left our solar system with many such objects. Once identified, these objects could be important enough that space agency’s send dedicated space missions to study them up close.

\section{Conclusions} \label{sec:Conclusions}
\medskip

Based on gravitational considerations of the Sun and the resulting fields of the Galaxy due to dark and baryonic matter, for the first time regions in space are estimated where a weak capture process leading to permanent weak capture would be occurring, in a simplified setting. Small openings into the solar Hill’s sphere has been determined to exist at about $3.81$ LY from the Sun in the direction of the galactic center or opposite to it. Permanent weak capture of interstellar objects into the Solar System is possible through these openings.  They would move chaotically within the  Hill's sphere to permanent capture about the Sun taking an arbitrarliy long time by infinitely many cycles. They would not collide with the Sun.   The permanent capture of interstellar comets and  rogue planets could occur. A rogue planet could perturb the orbits of the planets that may be possible to detect.

\appendix

\section{Supporting Calculations}
\label{sec:Appendix} 

\subsection{Kepler Energy,  $E_2$}
\label{subsec:KeplerEnergy}

In a $P_2$-centered inertial coordinate system,  $\bf{X} = (X_1, X_2)$,  the Kepler energy of $P$ relative to $P_2$ is
\begin{equation}
E_2 =  (1/2)|{\bm{\dot{X}}}|^2 - \mu |{\bf X}|^{-1},
\label{eq:KepEnergyInertial}
\end{equation}
where $X_1 = x_1 -1, X_2 = x_2$.  ${\bf{x}} = (x_1, x_2)$ are $P_1$-centered inertial coordinates.
\medskip

In $P_2$-centered rotating coordinates,  ${\bf{Y}} = (Y_1, Y_2)$, obtained by setting $Y_1 = y_1 - 1, Y_2= y_2$, where ${\bf{y}} = (y_1, y_2)$ are $P_1$-centered rotating coordinates defined
in (\ref{eq:DEs});
\begin{equation}
E_2({\bf Y}, {\bm \dot{Y}}) = (1/2) \dot{Y}^2 - \frac{\mu}{Y} -L({\bf Y}, {\bm \dot{Y}}) + (1/2) Y^2,
\label{eq:KepEnergyRot}
\end{equation}
where $L({\bf Y}, {\bm \dot{Y}}) = \dot{Y}_1Y_2 - \dot{Y}_2 Y_1$, $Y = |\bm{Y}|, \dot{Y} = |\bm{\dot{Y}}|$.   \\

\noindent
{\bf Acknowledgements}
\medskip

\noindent
 Many thanks to Princeton University, Department of Astrophysical Sciences.  E.B. was partially funded by NSF grant DMS-1814543.

\end{document}